\documentclass[osajnl,twocolumn,showpacs,superscriptaddress,10pt]{revtex4-1} 

\usepackage{amsmath,amssymb,graphicx}

\begin{document}

\title{Phase-Sensitive Noise Suppression in a Photoacoustic Sensor based on Acoustic Circular Membrane Modes}

\author{Mikael Lassen}\email{Corresponding author: ml@dfm.dk}

\author{Anders Brusch}

\author{David Balslev-Harder}

\author{Jan C. Petersen}

\affiliation{Danish Fundamental Metrology, Matematiktorvet 307, DK-2800 Kgs. Lyngby, Denmark}

\begin{abstract}
A photoacoustic (PA) sensor based on higher order acoustic modes is demonstrated. The PA sensor is designed to enhance the gas-detection performance and simultaneously suppress ambient noise sources (e.g. flow noise, electrical noise and external acoustic noise). Two microphones are used and positioned such that the PA signals are ($\pi$) out of phase. Ambient acoustic noise are approximately in the same phase and will be subtracted and thus improve the SNR. In addition, by placing the gas in- and outlets so that the gas flows through the node of the first higher order membrane mode the coupling of flow noise is approximately 20 dB lower compared with flow through the fundamental mode at 5 L/min. The noise reduction and thus the increase in sensitivity is demonstrated by measuring vibrational lines of methanol and methane using a broadband interband cascade laser emitting radiation at 3.38 $\mu$m. A signal-to-noise improvement of 20 (26 dB) using higher order modes are demonstrated compared with the fundamental mode. The maximum normalized noise equivalent absorption coefficient is 1.1 $\times 10^{-6}$ W cm$^{-1}$ Hz$^{1/2}$ for a flow of 5 L/min. It is anticipated that the noise cancelation strategy may find use in many practical industrial and environmental PA sensors, where ambient noise sources plays a crucial role for the absolute sensitive.
\end{abstract}



\maketitle              

\section{Introduction}

The development of sensitive trace gas sensors are of increasing importance in environmental, industrial and biological monitoring \cite{Hodgkinson2013}. Photoacoustic spectroscopy (PAS) is a very promising method due to its ease of use and its capability of allowing trace gas measurements at the sub-ppb level \cite{Harren2000,Miklos2001,Besson2006,Spagnolo2012}. The technique is based on the photoacoustic (PA) effect, where intensity modulated light is converted to sound in an absorbing material \cite{Rosencwaig1980Book,BEllPAS1881}. In a gas the sound waves are generated due to local heating via molecular collisions and de-excitation. A pressure sensitive device (e.g. microphone, tuning fork or cantilever) is used to monitor the generated acoustic waves \cite{Lindley2007,Patimisco2014,Koskinen2008}. The magnitude (in volts) of the microphone signal is given by: $S_{PA} = S_m P F \alpha$, where $P$ is the power of the incident radiation, $\alpha$ is the absorption coefficient, $S_m$ is the sensitivity of the microphone and $F$ is the cell-specific constant. From the equation it can be seen which parameters that needs to be optimized, to obtain a highly sensitive PA sensor. Ideally a highly sensitive PA system should amplify the PA sound wave only and reject acoustic and electrical noise as well as in-phase background signals from other materials in the PA cell. Unfortunately PA signals are generated by all absorbing materials in the PA cell and background signals can therefore originate from nonselective absorption in the cell windows and walls (in-phase noise). Ambient acoustic noise and flow noise may also couple to the PA signal, thus hinder sensitive measurements. In order to obtain an efficient noise suppression and improve the signal to noise ratio (SNR), many different designs of PA cells have been proposed and tested, aiming at various aspects of signal improvement, noise reduction, and ease of use \cite{Koskinen2008,lassen2014,Saarela2011,Manninen2012,Bernhardt2010,Dong2014,Lederman1986}.

\begin{figure}[h!]
\centerline{\includegraphics[width=.6\columnwidth]{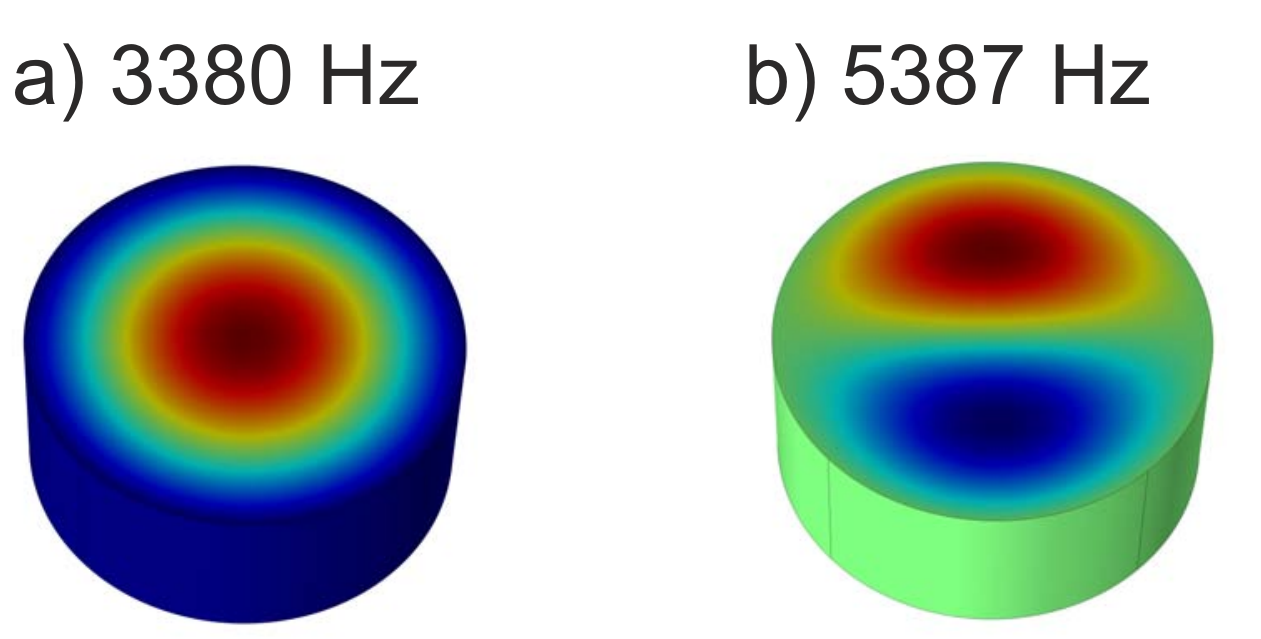}}
\caption{Simulations of the acoustic modes in a cylindrical PA cell. The acoustics pressure response of the first two eigenmodes are shown a) 3.38 kHz (0,1) and b) 5.39 kHz (1,1), respectively. The blue and red colors indicate the acoustic pressure with opposite phase. The modes are labelled $\nu_0$ and $\nu_1$, respectively.}
\label{comsolsimulations}
\end{figure}

In this paper it is demonstrated that the use of higher order acoustic modes increases the sensitivity by reducing the coupling of flow noise to the PA sensing devices and a further increase is shown by using two microphones in the same acoustic resonator (PA cell). The PA cell has a cylindrical symmetry resulting in Bessel function solutions to the wave equation. The microphones are mounted so that their responses are $\pi$ out of phase when the PA cell is resonant at the first higher order Bessel acoustic mode, typically denoted (1,1) while the fundamental mode is denoted (0,1). The modes under investigation are shown in Fig.~\ref{comsolsimulations}. By combining the microphone signals in antiphase $(S_{PA,1}-S_{PA,2})$, the PA signal amplitudes add up, while ambient acoustic noise, which are in approximately the same phase at the two microphones will be subtracted from each other and improve the SNR of the PA sensor. A number of PA schemes have already been investigated that takes advantages of phase-sensitive strategies \cite{Lederman1986,Bernhardt2010,Saarela2011,Kosterev2004}. Here a further improvement of the SNR is demonstrated by placing the gas in- and outlets so that the gas flows through the node of the first high order mode, thereby lowering the coupling of flow noise to the PA signal. The noise reduction, and thus the increase in sensitivity of the PA cell has been demonstrated by excitation of ro-vibrational lines of methane and methanol in the 3.38 $\mu$m wavelength region. The interest in these gasses is governed by the fact that methane and methanol are important trace gasses for environmental and biomedical sensing. Explicit an increase in the SNR of 26 dB was observed using the first higher order Bessel acoustic mode compared with the fundamental acoustic mode.

\section{Simulation of the acoustic response}

A typical setup for PAS involves an amplitude modulated light source and a resonant absorption cell with microphones, where the PA signal is enhanced by the acoustic resonances. The acoustics response of the PA cell is simulated using a finite element model (FEM) multi-physics simulation program (COMSOL). The pressure acoustic module is used to solve the wave equation. The boundary conditions were hard walls. The acoustic field is excited by applying a constant pressure at the walls. The cell material is made of PTFE (Teflon $\circledR$) and has a ring shaped geometry with a height of 25 mm and a diameter of 60 mm \cite{Manninen2012}. The two modes investigated and compared are the fundamental mode (0,1) and the first order Bessel mode (1,1). Note that the frequency of the (1,1) mode is 1.593 times the frequency of the (0,1) mode. The frequency response of the acoustic system is shown in Fig.~\ref{comsolsimulations}, where the first two eigenfrequencies are shown. The calculated eigenfrequencies are (a) 3.38 kHz and (b) 5.39 kHz, respectively, labeled $\nu_0$ and $\nu_1$, respectively. The simulations also allow the determination of the ideal positions of the microphones in the cell in order that they are $\pi$ out of phase when exciting the $\nu_1$ mode. The microphones are located at the midpoint of the lobes of the $\nu_1$ mode. Fig.~\ref{comsolsimulations}b) also shows that the node of $\nu_1$ mode is located diagonally at the center of the cell. The gas inlet and outlet are therefore located at this point in order to reduce the coupling of external acoustic noise to the $\nu_1$ mode. This has the consequence that flow noise will strongly couple to the fundamental $\nu_0$ mode. Since the flow is going directly through the $\nu_0$ mode where the acoustic amplitude is highest.

\section{Experimental Setup}

\begin{figure}[h!]
\centerline{\includegraphics[width=.8\columnwidth]{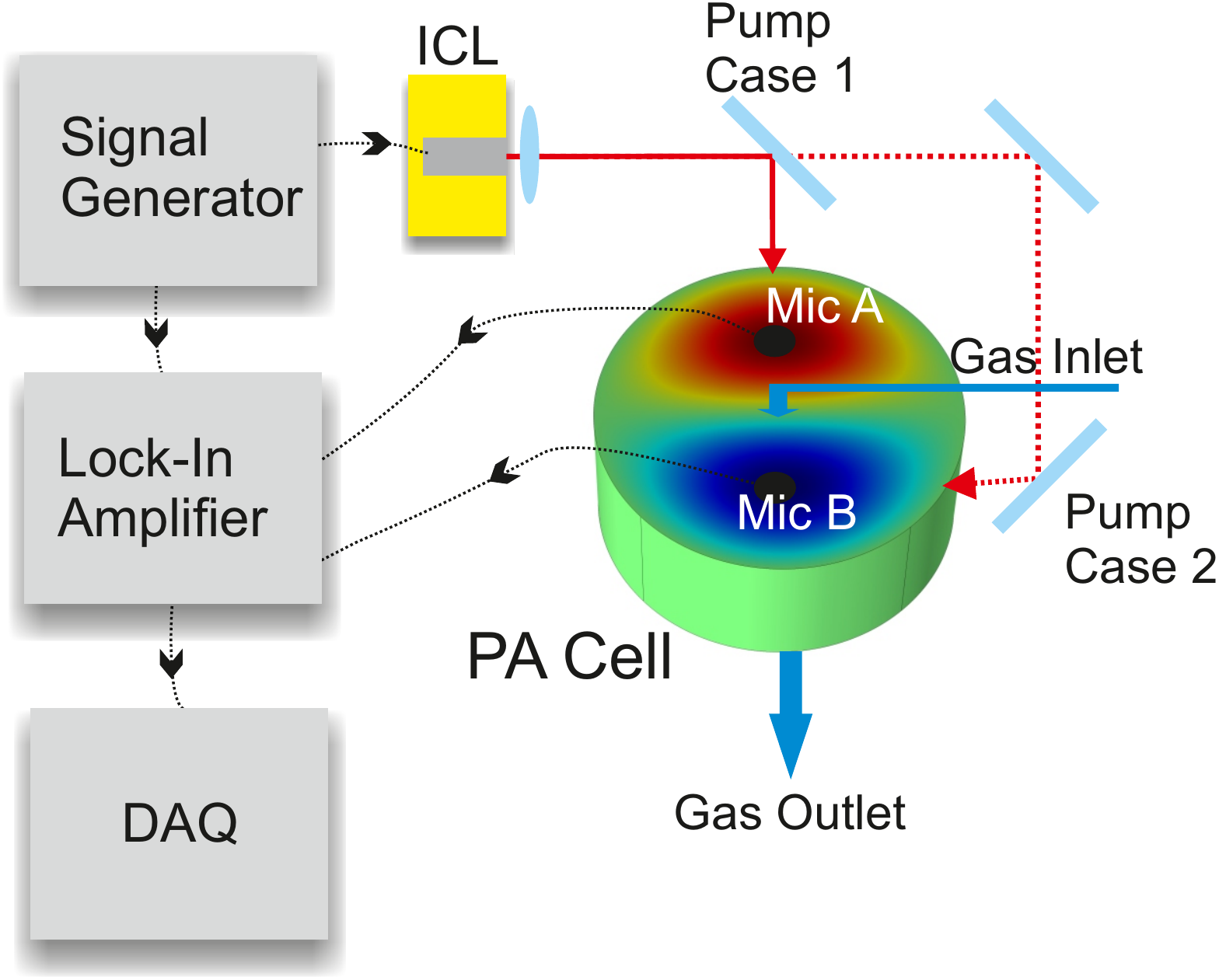}}
\caption{The experimental setup includes a 3.38 $\mu$m Interband Cascade Laser (ICL), two microphones attached to the PA cell, a signal generator for modulation of the laser current and a lock-in amplifier connected to a DAQ card.}
\label{experimentsetup}
\end{figure}

The schematics of the experimental setup is shown in Fig.~\ref{experimentsetup}. The laser source used is an Interband Cascade Laser (ICL) emitting radiation at 3.38 $\mu$m with a bandwidth of approximately 40 nm. The optical output power is approximately 60 mW. Just in front of the laser output an AR coated aspheric lens (F = 1.873 mm and NA = 0.85) is placed. The lens collimates the laser beam that enters the PA cell via an uncoated 3 mm thick Zinc Selenide  window. The optical transmission is approximately 70$\%$ at 3.38 $\mu$m and the absorption coefficient is approximately 10$^{-3}$ cm$^{-1}$. The uncoated windows give rise to a small in-phase background signal. The laser beam then hits the cell wall opposite of the window and is scattered. The PA cell is made of PTFE and has a cylindrical shape. The radius of the cell is 30 mm and the hight is 25 mm. PTFE material is used for several reasons. Firstly, PTFE has an average reflectivity of $>90\%$ at 3.38 $\mu$m. Secondly, the PTFE walls can decouple the in-phase background absorption signal from the PA signal due to a low-passing effect of the modulated light. Finally, the background signal does not contribute to the PA signal due to the uniform distribution of the scattered light inside the cell \cite{lassen2014}. Due to these effects it is anticipated that background signals only arise from the cell windows. Two microphones are positioned to be $\pi$ out of phase when the PA cell is resonant at the $\nu_1$ mode. Subtracting the microphone signals results in the PA signal amplitudes add up, while ambient acoustic noise, which are in approximately the same phase at the two microphones, will be subtracted from one another and thus improve the SNR. The signals from the microphones were amplified and filtered with a 7 kHz bandpass filter (3-10 kHz) before further signal processing. The optical radiation is amplitude modulated using a square wave modulation on the the laser current, controlled by a signal generator. This generates a PA signal at the modulation frequency. The data is processed using a fast Fourier transformation (FFT) analyse and a lock-in amplifier. The ICL modulation is controlled by a signal generator, which also acts as the local oscillator for the lock-in amplifier. The peak-to-peak modulation is 60 mW, and approximately 40 mW is coupled into the PA cell. The data from the lock-in amplifier is collected with a 250 kS/s DAQ card with 16 bit resolution.

\section{Experimental frequency response and coupling of flow noise}

The experiments were performed by excitation of molecular ro-vibrational modes of methane and methanol in the 3.38 $\mu$m wavelength region.

\begin{figure}[h!]
\centerline{\includegraphics[width=.9\columnwidth]{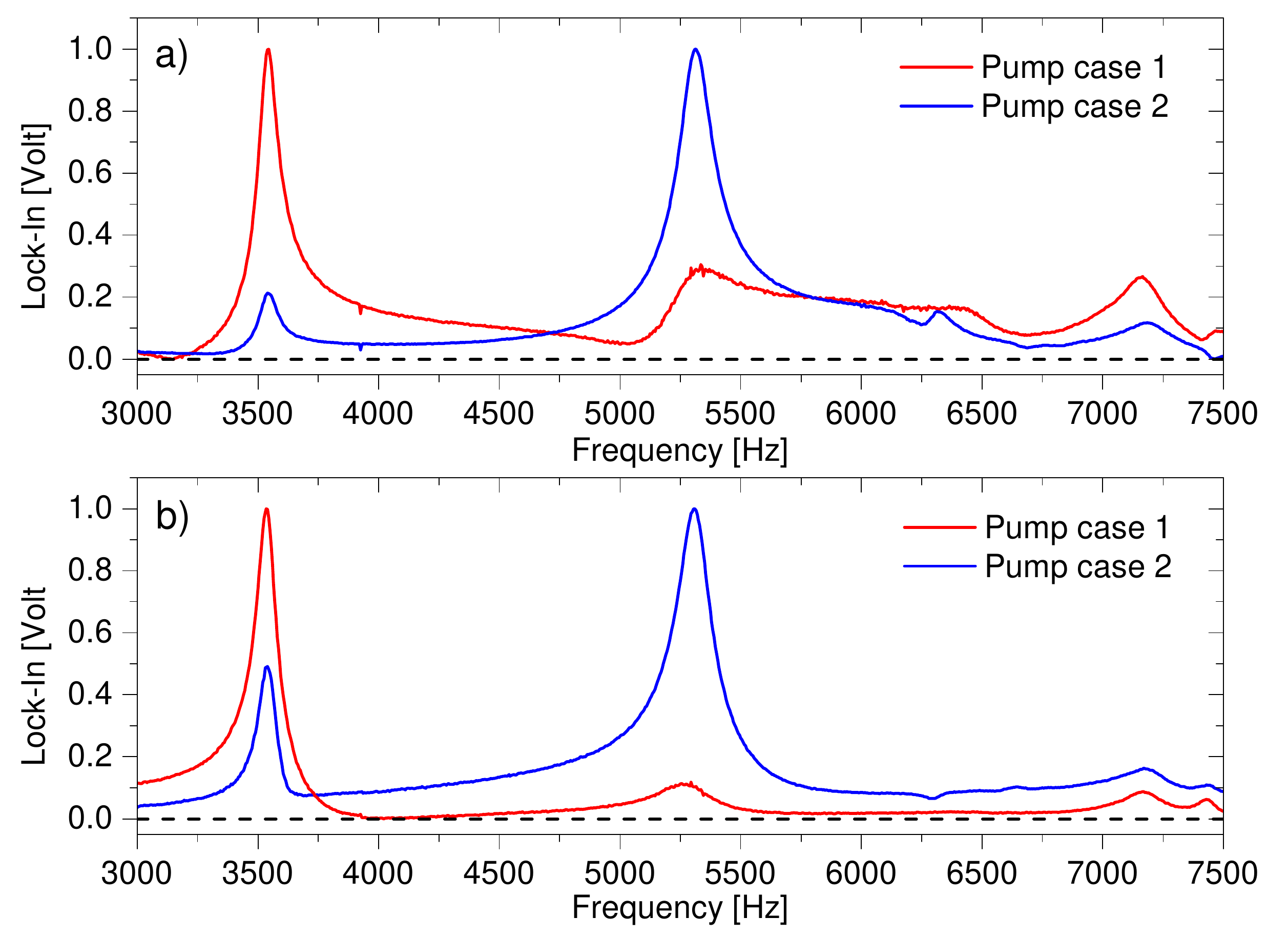}}
\caption{The normalized PA response of the PA cell for the two microphones and for the two different pump schemes. a) Response of Mic A and b) response of Mic B. Red curves are for pump scheme 1 and blue curves are for pump scheme 2. The PA signal is generated using a unknown concentration of methanol. The data is processed with a SRS Lock-In amplifier with 1 second integration time.}
\label{pumpcases}
\end{figure}

In the first experiment the overlap between the light field and the acoustics mode were investigated. The measurements are performed on methanol vapor with a unknown concentration. The conversion of energy from the light field to the acoustics mode is given by the overlap integral, $\int p_j I(\omega) dV$, where $p_j$ is the acoustic pressure field and $I(r,\omega)$ is the optical intensity distribution in the PA cell \cite{Rosencwaig1980Book}. Two pump schemes were used as illustrated in Fig.~\ref{experimentsetup}. Scheme 1, the PA cell is pumped through both lobes of the $\nu_1$ mode and through the middle of the $\nu_0$ mode. In this pumping scheme the calculated overlap integral results in a energy conversion to the $\nu_1$ mode of zero. Scheme 2, the PA cell is pumped through one of the lobes of the $\nu_1$ mode. The ratio of the overlap integral between the two modes $\nu_1$ and $\nu_0$ is calculated to be approximately 3. This means that the $\nu_1$ amplitude should be approximately 3 times larger than the $\nu_0$ amplitude. Fig.~\ref{pumpcases} depict the experimental data for for the two pump schemes and the response from the two microphones Fig.~\ref{pumpcases}(a) and Fig.~\ref{pumpcases}(b), respectively. It is clearly observed that depending on which pump scheme is used the energy is transferred mostly to the $\nu_0$ (red curves) or the $\nu_1$ (blue curves), respectively. For both pump schemes the laser beam hits the cell wall opposite of the window and is scattered so that the light field is uniform distributed over all angles. The background signal does therefore not contributed to the PA signal since an uniform distribution of the light field can only contribute with a non-resonant contribution to the PA signal \cite{Rosencwaig1980Book,lassen2014}. This can be seen from Fig.~\ref{pumpcases} since if the scattered uniform light field would contribute to the acoustic resonances. Pump scheme 1 and 2 would result in approximately the same overlap integral between the light field and the acoustics modes, thus the acoustic response would be identical in the two different pump schemes. The experimental ratios between the $\nu_1$ and $\nu_0$ agree well with the calculated ratios. It is noticed that the laser beam in scheme 1 is not perfect aligned through the center of the PA cell and therefore energy is converted to the $\nu_1$ mode. From Fig.~\ref{pumpcases} it can also be seen that the acoustic response of the PA cell is in very good agreement with the simulations. The experimental eigenfrequencies of the first two Bessel modes are at (a) 3.53 kHz and (b) 5.29 kHz, respectively. This results in a ratio of 1.5 between the frequencies of the two modes, which is in good agreement with the theory. Fig.~\ref{pumpcases} shows that the Q-factor of the two modes are approximately 28.

\begin{figure}[h!]
\centerline{\includegraphics[width=.8\columnwidth]{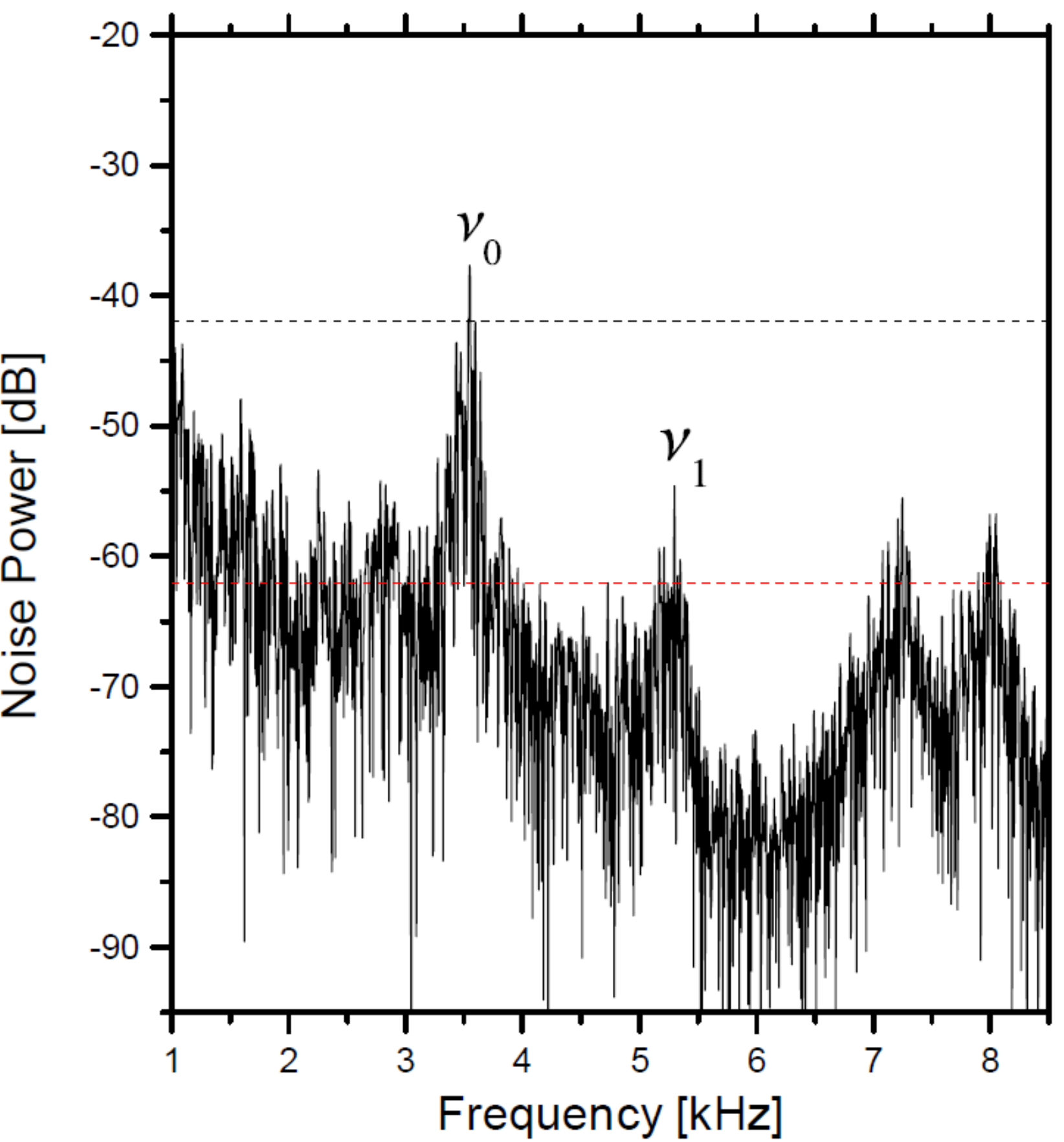}}
\caption{The coupling of flow noise to the acoustic modes $\nu_0$ and $\nu_1$. Noise spectrum (FFT) of PA cell with a 5L/min flow. $\nu_0$ is approximately 20 dB more affected by the flow than the $\nu_1$. }
\label{flownoise}
\end{figure}

The comparison of flow noise for the two modes are shown in Fig.~\ref{flownoise}. It can be seen that the $\nu_0$ mode is approximately 10 (20 dB) more affected by flow noise than the $\nu_1$ mode as expected from the above simulation. Using the $\nu_0$ mode as the PA frequency has therefore the consequence that the sensitivity will be at least one order of magnitude lower than using the $\nu_1$ mode due to the flow noise.

\section{Two microphone noise cancelation strategy}

\begin{figure}[h!]
\centerline{\includegraphics[width=.8\columnwidth]{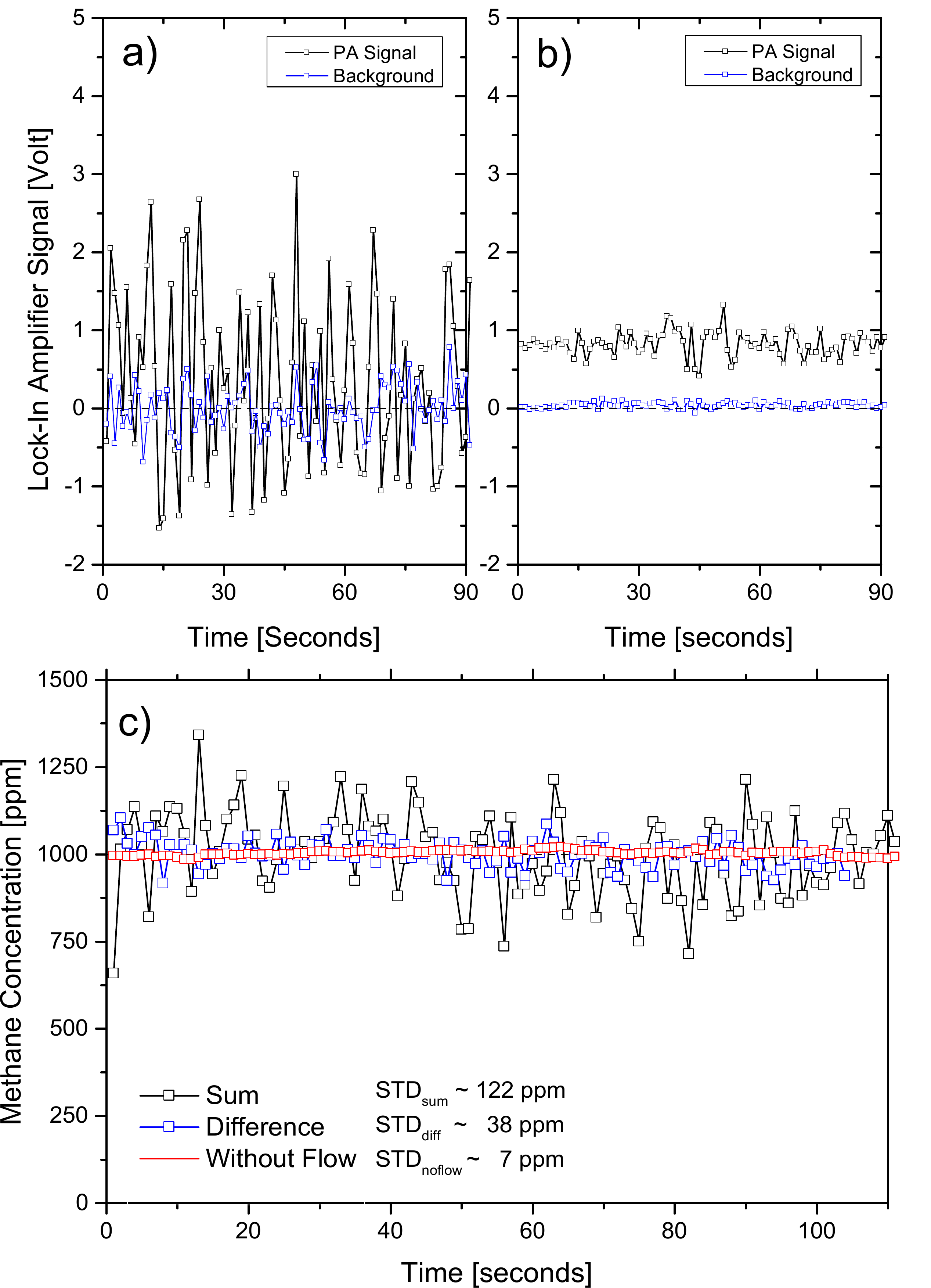}}
\caption{PA measurements of methane with a flow rate of 5 L/min. a) Excited PA signal at 3.53 kHz. b) Excited PA signal at 5.29 kHz. The PA signals are processed with a lock-in amplifier with 300 $\mu$s integration time. c) Sum (black curve) and difference (blue curve) signal for a PA signal at 5.29 kHz. The red curve is the PA signal without flow.  The data is processed with a lock-in amplifier with 3 seconds integration time.}
\label{sensitivity}
\end{figure}

In order to enhance the SNR further the two microphones are placed in each lobe of the $\nu_1$ mode. Combining the microphone signals in anti-phase, the PA signal amplitudes add up for the $\nu_1$ mode, while ambient acoustic noise, which are in approximately the same phase at the two microphones, will be subtracted from each other and thus improve the SNR of the PA sensor. In order to quantify the increase in sensitivity of the PA sensor the PA cell is flowed with a 1000 ppm methane in N$_2$ mixture at 5 L/min. The generated PA signal at the $\nu_0$  and the $\nu_1$ modes is compared in Fig.~\ref{sensitivity}. The PA signal from the two microphones are amplifier and then added/subtracted before processed with a lock-in amplified. Fig.~\ref{sensitivity}(a) and (b) clearly indicate that the SNR of the PA signal is enhanced. Fig.~\ref{sensitivity}(a) shows the added microphone signals for a PA signal at 3.53 kHz with a SNR of 0.3 and Fig.~\ref{sensitivity}(b) shows the subtracted microphone signals for a PA signal at 5.29 kHz with a SNR of 6. The relative increase in SNR is therefore approximately 20 (26 dB) for the $\nu_1$ mode compared with the $\nu_0$ mode. This increase is due to the lower coupling of flow noise and the two microphone noise cancelation strategy. Fig.~\ref{sensitivity}(c) shows the sum and difference between the two microphones at a modulation frequency 5.29 kHz and with the integration time of the locking-in amplifier set to 3 seconds. The sum signal has a standard deviation (STD) of 122 ppm and the difference signal a STD of 38 ppm. This corresponds to a normalized noise-equivalent absorption of 1.1 $\times 10^{-6}$ W cm$^{-1}$ Hz$^{1/2}$ and 3.6 $\times 10^{-5}$ W cm$^{-1}$ Hz$^{1/2}$ for the difference and sum signal, respectively. This confirms that the two microphone measurement strategy indeed increases the SNR and thus the sensitivity of the system. Note that without flow the achieved STD of the system is 7 ppm in 3 seconds measured at a modulation frequency of 5.29 kHz. This corresponds to a normalized noise-equivalent absorption of 6.3 $\times 10^{-8}$ W cm$^{-1}$ Hz$^{1/2}$. It is anticipated that the sensitivity can be further enhanced by improving the acoustic insulation, use more sensitive microphones, optimized electronics and improved temperature control of the ICL laser. This will make the sensor comparable with state of the art PA sensors \cite{Patimisco2014}.

\section{Conclusion}

It was shown that by proper design and by including an additional microphone the sensitivity of a PA sensor can be enhanced leading to improved gas-detection performance and simultaneously suppression of ambient noise sources (e.g. flow noise, electrical noise and external sounds). The microphones are placed such that the PA signals are out of phase, while ambient acoustic noise measured with the two microphones have approximately the same phase. Thus by  subtracting from each other the SNR of the PA sensor will be increased. The noise reduction and thus the increase in sensitivity is demonstrated by measuring ro-vibrational lines of methane and methanol in the 3.38 $\mu$m wavelength region. A SNR improvement of 26 dB of the subtracted signal at the $\nu_1$ mode compared to the added signal  at the $\nu_0$ mode has been demonstrated. The PA cell was made of PTFE and it was shown that the PTFE walls can decouple the in-phase background absorption signal from the PA signal due to a low-passing effect of the modulated light. It is anticipated that the demonstrated method of combining the microphone signals in anti-phase which are acoustically out of phase may find use as a noise compensation strategy in many practical industrial and environmental PA sensors, where ambient noise sources plays a crucial role for the absolute sensitive.


\end{document}